# Leakage Mitigation in Heterodyne FMCW Radar For Small Drone Detection with Stationary Point Concentration Technique

Junhyeong Park, *Student Member, IEEE*, Seungwoon Park, Do-Hoon Kim, and Seong-Ook Park, *Senior Member, IEEE*

***Abstract*— Recently, as the drones have become smaller and smarter, they are emerging as new threats to the military and the public. To prevent the threats, the development of the radar to detect the small drones is needed. The frequency modulated continuous wave (FMCW) radar is one of the radar types for the drone detection. The heterodyne architecture for the FMCW radar is often applied to resolve the dc offset problem. However, the leakage from a transmitter into a receiver, the notorious problem of the FMCW radar, is still a challenging problem. Especially, the phase noise of the leakage increases the noise floor and deteriorates the signal to noise ratio. In order to mitigate this problem, in this paper, the stationary point concentration technique is proposed. Without additional hardware, the proposed technique can be implemented through frequency planning and digital signal processing. The results show that the proposed technique significantly lowers the noise floor over the desired range domain.**

## I. Introduction

THE drone market is growing faster and faster. Nowadays, in the public, the drones are accepted as a new means of the filming, the racing sport, and the future vehicle for the delivery service. However, they also bring many threats such as crash and terrorism. Even now, the drones are getting smaller, faster, smarter, and easier to control, thus the fear of the threats are also growing. To prevent these threats, development of a drone detection radar is inevitable. One of the typical radar types for the drone detection is the frequency modulated continuous wave (FMCW) radar [1]-[5]. In comparison with the pulse radar, the FMCW radar has advantages in the cost, the peak power, and the minimum detectable range. Particularly, the limitation of the minimum detectable range is unavoidable in the pulse radar since the receiver is turned off while the pulse is being emitted [6]. On the other hand, the FMCW radar continuously receives the electromagnetic wave. Therefore, the operational mechanism of the FMCW gives an advantage in the target detection at the near distance.

There are two architectures for the FMCW radar. One is homodyne, and the other is heterodyne. The homodyne FMCW radar directly mixes the reference FMCW signal with received FMCW signals. Thus, resulting beat signals are generated in the baseband. However, due to the imperfect isolation between the local oscillator (LO) port and the radio frequency (RF) port of the mixer in the RF stage, and between the mixer and the low noise amplifier (LNA), the self-mixing phenomenon occurs. The self-mixing produces undesired dc component that is called as dc offset, and it can saturate the following stages such as amplifier and analog to digital converter (ADC) and corrupt the signals in the baseband [7]. On the other hand, the heterodyne FMCW radar suffers much less from the dc offset problem, because it produces beat signals in the intermediate frequency (IF) stage. The IF stage makes it easier to attenuate the dc offset through a high pass or a band pass filter. The heterodyne FMCW radar has been chosen to take this advantage in many published papers [4], [8]-[11].

Although the heterodyne FMCW radar has many advantages, solving the leakage from the transmitter (TX) into the receiver (RX) that is the notorious issue of the FMCW radar is not one of them. In the monostatic FMCW radar, the leakage occurs in a circulator due to its inefficient isolation capability, and the mismatch in antenna also leads to the leakage [12]-[14]. In case of the bistatic FMCW radar, the transmitted signal is leaked into the receiver by the mutual coupling between the TX and the RX antennas, and objects surrounding the antennas can also be the cause of the leakage [15]. The leakage produces severe problems. Because its power is typically much higher than that of the returned signals, the LNA can be saturated [12], [15]. Additionally, the dynamic range of the FMCW radar is limited by the phase noise of the leakage [12].

In detecting small drones, the most basic and important goal is to increase the signal to noise ratio (SNR), and this can be done by resolving the leakage problem. There have been many studies to resolve the leakage problem in the FMCW radar. In

This work has been submitted to the IEEE for possible publication. Copyright may be transferred without notice, after which this version may no longer be accessible. This research was supported by a grant (18ATRP-C108186-04) from UAV Safety Technology Research Program funded by the Ministry of Land, Infrastructure and Transport of Korean Government.

The authors are with the School of Electrical Engineering, Korea Advanced Institute of Science and Technology, Daejeon 34141, South Korea (e-mail: bdsfh0820@kaist.ac.kr; physicsoly@kaist.ac.kr; dohoonh@kaist.ac.kr; soparky@kaist.ac.kr).

[4], [8], there were attempts to reduce the leakage by putting the TX apart from the RX. To increase the distance between the TX and the RX, 106.2 m fiber-optic cables that have low-loss characteristic were used in [4]. In case of [8], 75 Ω coaxial cables whose lengths are 10 m for TX and 30 m for RX were used. In [13], adding a closed loop leakage canceller for the monostatic radar was proposed. The closed loop leakage canceller adaptively generates an error vector which includes amplitude and phase information of the leakage. In [12], [14], the scheme in [13] was implemented through the digital signal processing (DSP) approach and improved by up-converting the error signal to a preselected reference frequency to overcome the dc offset problem. In [16]-[19], the reflections caused by fixed objects in front of the antennas were studied. These reflections were introduced as short-range (SR) leakage. Based on the correlation statistics of the decorrelated phase noise (DPN), delay time can be selected, and it was implemented by adding an artificial on-chip target in radio frequency (RF) stage. Then, after mixing expected parameters in digital intermediate frequency (IF) stage, a sampled IF signal which is similar to the sampled IF signal of the SR leakage could be generated. The cancellation of the SR leakage was attempted by subtracting the two IF signals from each other.

We have focused on the dominant leakage and its phase noise. In the recent conference paper, we proposed a new down-conversion concept to mitigate the leakage [20]. However, because the hardware system was not complete at that time, there were only simple simulation results, which are not enough to demonstrate the actual effectiveness of the proposed concept. Also, more analyses based on both the theory and simulation to verify the proposed concept were required. In this paper, we provide detailed analyses on the proposed theory with the simulations. In addition, we introduce detailed procedures to implement the proposed concept in practice. The performance of the proposed technique is predicted with the simulations, and demonstrated by the experiment results.

So far, the trend in the research field on the leakage of the FMCW radar has been cancelling out the leakage by trying to create the same signal with the leakage and subtracting it from the received signal [12]-[14], [16]-[19]. Instead of this trend, the proposed concept suggests a novel approach to significantly mitigate the leakage. The proposed concept extracts the frequency and the constant phase information of the beat signal of the leakage in the digital IF domain. After generating a digital numerically controlled oscillator (NCO) which has the extracted frequency and constant phase values of the beat signal of the leakage, the received signal is finally down-converted with the digital NCO. By doing this, the frequency and the constant phase in the beat signal of the leakage are removed, so the term of the major cause of the phase noise is removed, and the phase noise of the leakage is concentrated on the stationary point of the cosine function. Therefore, the magnitude of the leakage phase noise is significantly attenuated, and the noise floor is decreased. Unlike the previous techniques, the proposed technique can be implemented through the frequency planning and DSP without additional hardware parts. In order to emphasize the key point of the proposed technique, we named it the stationary point concentration (SPC) technique. For the verification, the results of two experiments are shown. First, we show how much the noise floor is decreased in a situation that only the leakage exists without any target. Second, with the small drones, DJI Inspire 1 and DJI Spark, we demonstrate the increased SNR. MATLAB R2017a is used for all DSP in the simulations and the experiments.

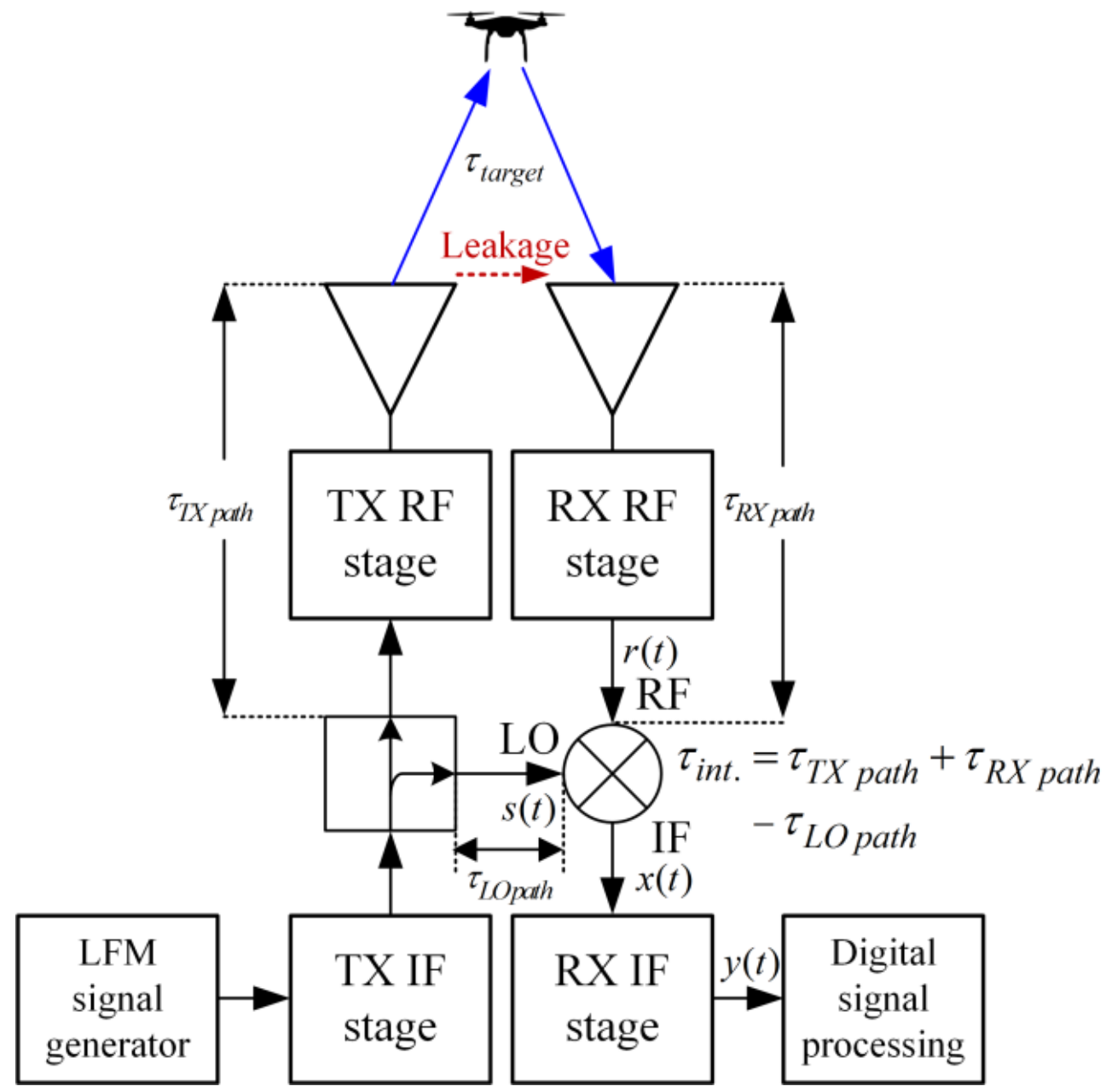

Fig. 1. Block diagram of the heterodyne FMCW radar.

In Section II, the theories of the heterodyne FMCW radar and the proposed SPC technique are introduced. In Section III, detailed procedures of the SPC technique are explained. In Section IV, analysis on the SPC technique and the prediction of the experiment results are carried out with the simulations. Then, in Section V, the FMCW radar that we used for the experiments are presented, and the two experiments for the verification of the SPC technique are explained in Section VI. The experiment results and the discussion are presented in Section VII, followed by the conclusion in Section VIII.

## II. Theories

### A. Heterodyne FMCW Radar

Fig. 1 shows the block diagram of the heterodyne FMCW radar. A linear frequency modulated (LFM) signal, which is also called as ramp signal or chirp signal is divided into two paths by a splitter. One passes the TX RF stage which includes cables, LO, mixers, filters, isolators, and power amplifiers (PAs). The LFM signal is up-converted to the RF band by being mixed with the LO signal, then it is radiated through the TX antenna. The other LFM signal on the other path, called as reference LFM signal, is mixed with the received LFM signals. With this FMCW mixing process, the beat signals are extracted and these include distance, and even Doppler information of targets.

In Fig. 1, the reference LFM signal at the LO port, $s(t)$, can be defined as follows:

$$s(t) = A_S \cos\left(2\pi f_{TX} t + \pi\alpha t^2 + \theta_S + \varphi_S(t)\right) \tag{1}$$

for $0 < t < T$, where $A_S$ and $f_{TX}$ are the amplitude and the start frequency of the reference LFM signal, $\alpha = BW/T$ is the slope of the chirp, $BW$ and $T$ are the sweep bandwidth and the sweep period, $\theta_S$ and $\varphi_S(t)$ are the constant phase and the phase noise of the reference LFM signal. The delay from the splitter to the mixer for the FMCW mixing, i.e., $\tau_{LO\ path}$, is considered together with the other delays at the RF port.

Passing the TX RF stage, the LFM signal is delayed by $\tau_{TX\ path.}$. After being transmitted from the TX antenna, the leaked LFM signal directly enters the RX antenna. Then, the signals reflected by the targets follow. After being received through the RX antenna, these signals pass the RX RF stage which includes low noise amplifiers (LNAs), isolators, LO mixers, filters, and cables, therefore another internal delay, $\tau_{RX\ path}$, is added. In this paper, only the dominant leakage is considered, and we assume that the time delay due to the spatial path between the antennas is negligible in the quasi-monostatic radar. Therefore, the received signals at the RF port, $r(t)$, can be expressed as follows:

$$\begin{aligned} r(t) = & \underbrace{A_L \cos\left(2\pi f_{RX}(t-\tau_{int.}) + \pi\alpha(t-\tau_{int.})^2 + \theta_R + \varphi_L(t)\right)}_{Leakage} \\ & \underbrace{+\sum_{k=1}^{K} A_{T,k} \cos\left(2\pi f_{RX}(t-\tau_{int.}-\tau_{T,k}) + \pi\alpha(t-\tau_{int.}-\tau_{T,k})^2 + \theta_R + \varphi_{T,k}(t)\right),}_{Targets} \end{aligned} \tag{2}$$

where $\theta_R$ is the constant phase of the received LMF signals, $A_L$ & $A_{T,k}$ and $\varphi_L(t)$ & $\varphi_{T,k}(t)$ are the amplitudes and the phase noises of the leakage and the target LFM signals at the RF port respectively, $f_{RX}$ is start frequency at the RF port, $\tau_{int.}$ is total internal delay, $\tau_{int.} = \tau_{TX\ path} + \tau_{RX\ path} - \tau_{LO\ path}$, and $\tau_{T,k}$ is the round-trip delay to the targets.

In the heterodyne architecture, there are LO signals in the TX RF stage and the RX RF stage for the up-conversion and the down-conversion. These LO signals have their own phase noise, $\varphi_{TX\ RF\ LO}(t)$ and $\varphi_{RX\ RF\ LO}(t)$, which are different from that of the LFM signals, because the voltage controlled oscillators (VCO) in the phase locked loops (PLLs) for the LO signals are independent. $\varphi_L(t)$ and $\varphi_{T,k}(t)$ can be represented as follows:

$$\begin{aligned} \varphi_L(t) \approx & \varphi_S(t-\tau_{int.}) + \varphi_{TX\ RF\ LO}(t-\tau_{TX\ path}-\tau_{RX\ path}) \\ & -\varphi_{RX\ RF\ LO}(t-\tau_{RX\ path}) \end{aligned} \tag{3}$$

$$\begin{aligned} \varphi_{T,k}(t) \approx & \varphi_S(t-\tau_{int.}-\tau_{T,k}) \\ & +\varphi_{TX\ RF\ LO}(t-\tau_{TX\ path}-\tau_{RX\ path}-\tau_{T,k}) \\ & -\varphi_{RX\ RF\ LO}(t-\tau_{RX\ path}-\tau_{T,k}). \end{aligned} \tag{4}$$

Finally, after the FMCW mixing, if we consider only the desired terms that are IF beat signals, these terms at the IF port of the mixer, $x(t)$, can be written as

$$\begin{aligned} x(t) = & x_{IF\ leakage}(t) + x_{IF\ targets}(t) \\ = & \frac{A_S A_L}{2} \cos\left( 2\pi(\overbrace{\underbrace{f_{TX}-f_{RX}}_{f_{IF\ carrier}} + \underbrace{\alpha\tau_{int.}}_{f_{beat\ leakage}}}^{f_{IF\ beat\ leakage}})t + \underbrace{\theta_S + 2\pi f_{RX}\tau_{int.} - \pi\alpha\tau_{int.}^2 - \theta_R}_{\theta_{IF\ leakage}} + \varphi_{IF\ leakage}(t) \right) \\ & +\sum_{k=1}^{K} \frac{A_S A_{T,k}}{2} \cos\left( 2\pi(\overbrace{\overbrace{\underbrace{f_{TX}-f_{RX}}_{f_{IF\ carrier}} + \underbrace{\alpha\tau_{int.}}_{f_{beat\ leakage}}}^{f_{IF\ beat\ leakage}} + \underbrace{\alpha\tau_{T,k}}_{f_{beat\ targets,k}}}^{f_{IF\ beat\ targets,k}})t \right. \\ & \left. + \underbrace{\theta_S + 2\pi f_{RX}(\tau_{int.}+\tau_{T,k}) - \pi\alpha(\tau_{int.}+\tau_{T,k})^2 - \theta_R}_{\theta_{IF\ targets,k}} + \varphi_{IF\ targets,k}(t) \right). \end{aligned} \tag{5}$$

The phase noises of the IF beat signals of the leakage and the targets, $\varphi_{IF\ leakage}(t)$ and $\varphi_{IF\ targets,k}(t)$, are

$$\begin{aligned} \varphi_{IF\ leakage}(t) = & \underbrace{\varphi_S(t) - \varphi_S(t-\tau_{int.})}_{DPN} \\ & -\varphi_{TX\ RF\ LO}(t-\tau_{TX\ path}-\tau_{RX\ path}) \\ & +\varphi_{RX\ RF\ LO}(t-\tau_{RX\ path}). \end{aligned} \tag{6}$$

$$\begin{aligned} \varphi_{IF\ targets,k}(t) = & \underbrace{\varphi_S(t) - \varphi_S(t-\tau_{int.}-\tau_{T,k})}_{DPN} \\ & -\varphi_{TX\ RF\ LO}(t-\tau_{TX\ path}-\tau_{RX\ path}-\tau_{T,k}) \\ & +\varphi_{RX\ RF\ LO}(t-\tau_{RX\ path}-\tau_{T,k}). \end{aligned} \tag{7}$$

In (5), we assume $f_{TX} > f_{RX}$ so that the beat frequency is added to the IF carrier frequency, $f_{IF\ carrier}$. If $f_{TX} < f_{RX}$, the beat frequency is subtracted from the IF carrier frequency, $|f_{TX} - f_{RX}|$. Also, note that the beat frequency of the leakage, $f_{beat\ leakage}$, comes from the total internal delay.

In (6) and (7), there are many phase noise terms. In the power spectral density (PSD), the total phase noises follow the larger phase noise. The terms, $\varphi_S(t)-\varphi_S(t-\tau_{int.})$ and $\varphi_S(t)-\varphi_S(t-\tau_{int.})-\tau_{T,k}$, which are also called as DPN, have the range correlation effect [16]-[19], [21]. Because the range correlation effect reduces the magnitude of $\varphi_S(t)$ [16], [21], the phase noises of the LOs in the RF stages are dominant over the DPN terms.

After the IF beat signals pass some blocks in the RX IF stage, the final down-conversion to get rid of the IF carrier frequency is carried out, and only the beat signals are extracted. The IF carrier frequency, $f_{IF\ carrier}=f_{TX}-f_{RX}$, is decided through the frequency planning of the radar, so it is a known value. Thus, $f_{IF\ carrier}$ is chosen as the frequency of the last LO, that is

$$LO_{common}(t)=A_{LO}\cos\left(2\pi f_{IF\ carrier}t+\theta_{LO}+\varphi_{LO}(t)\right), \quad (8)$$

where $A_{LO}$, $\theta_{LO}$, and $\varphi_{LO}(t)$ are the amplitude, the constant phase, and the phase noise of the last LO, respectively. If (5) is mixed with (8) and passes a low pass filter, then the final signal, $y(t)$, in the heterodyne FMCW radar can be represented as follows:

$$\begin{aligned}y(t)&=y_{leakage}(t)+y_{target}(t)\\&=\underbrace{A_{leakage}\cos\left(2\pi f_{beat\ leakage}t+\theta_{leakage}+\varphi_{leakage}(t)\right)}_{Leakage}\\&\quad\underbrace{+\sum_{k=1}^{K}A_{targets,k}\cos\left(2\pi\left(f_{beat\ leakage}+f_{beat\ targets,k}\right)t+\theta_{targets,k}+\varphi_{targets,k}(t)\right).}_{Targets}\end{aligned} \quad (9)$$

In (9), $A_{leakage}=A_SA_LA_{LO}/4$, $A_{targets,k}=A_SA_{T,k}A_{LO}/4$, $\theta_{leakage}=\theta_{IF\ leakage}-\theta_{LO}$, $\theta_{targets,k}=\theta_{IF\ targets,k}-\theta_{LO}$, $\varphi_{leakage}(t)=\varphi_{IF\ leakage}(t)-\varphi_{LO}(t)$, and $\varphi_{targets,k}(t)=\varphi_{IF\ targets,k}(t)-\varphi_{LO}(t)$. Then, according to the cosine sum identity, $y_{leakage}(t)$ in (9) can be transformed as follows:

$$\begin{aligned}y_{leakage}(t)&=A_{leakage}\cos\left(2\pi f_{beat\ leakage}t+\theta_{leakage}\right)\cos\left(\varphi_{leakage}(t)\right)\\&\quad-A_{leakage}\sin\left(2\pi f_{beat\ leakage}t+\theta_{leakage}\right)\sin\left(\varphi_{leakage}(t)\right).\end{aligned} \quad (10)$$

Generally, since the phase noise is much smaller than 1 [16], [21], [22], (10) can be approximated as follows:

$$\begin{aligned}y_{leakage}(t)&\approx A_{leakage}\cos\left(2\pi f_{beat\ leakage}t+\theta_{leakage}\right)\\&\quad\underbrace{-A_{leakage}\varphi_{leakage}(t)\sin\left(2\pi f_{beat\ leakage}t+\theta_{leakage}\right)}_{Major\ cause\ of\ the\ noise\ floor\ rise}.\end{aligned} \quad (11)$$

As expressed in (11), the phase noise of the leakage signal is up-converted to $f_{beat\ leakage}$, and it manifests itself as voltage or current noise. Also, $A_{leakage}$ is generally much larger than $A_{targets,k}$ [12], [15]. These properties create a strong leakage signal in the power spectrum and it leads to the increase in the overall noise floor.

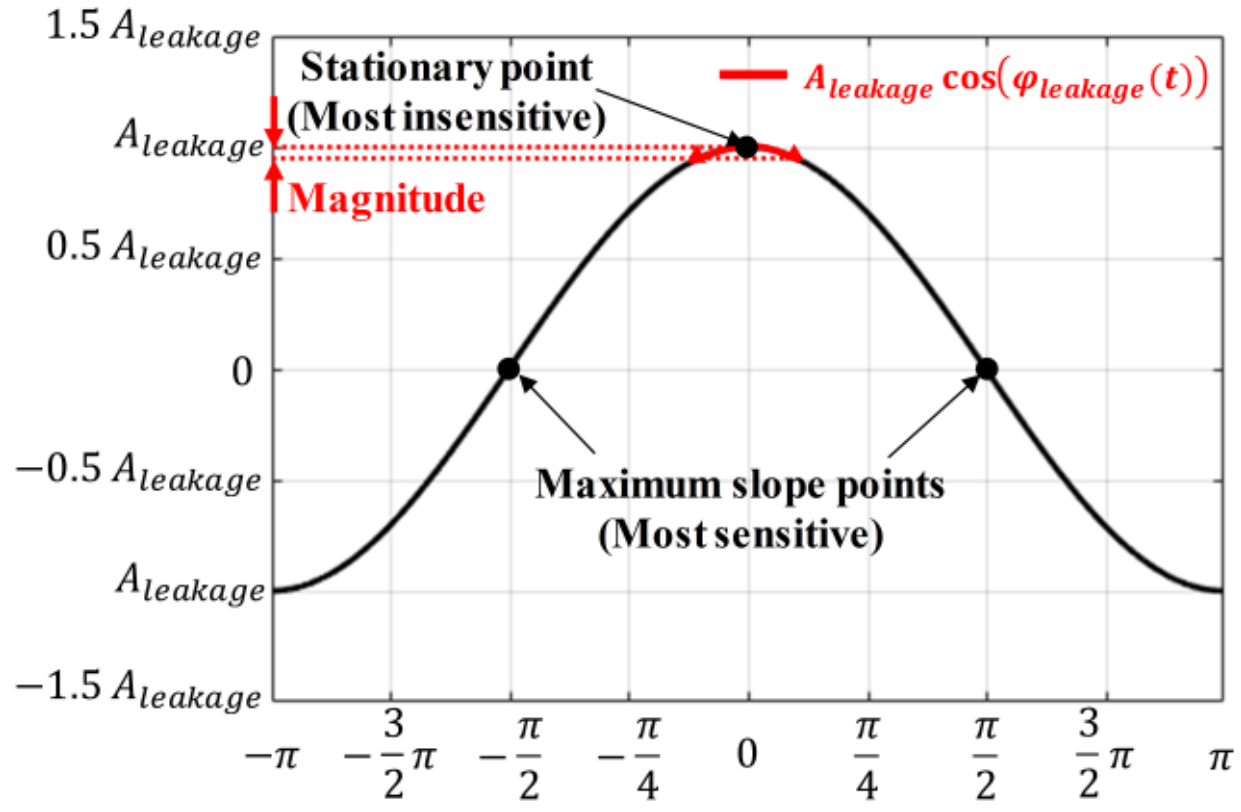

Fig. 2. Phase noise of the leakage on the stationary point.

## B. Stationary Point Concentration Technique

The proposed SPC technique is applied after the FMCW mixing process. When the final down-conversion is conducted, the SPC technique uses the exact IF beat frequency of the leakage, $f_{IF\ beat\ leakage}=f_{IF\ carrier}+f_{beat\ leakage}$, and the exact constant phase, $\theta_{IF\ leakage}$ for the last LO, whereas the common method just uses the known IF carrier frequency, $f_{IF\ carrier}$. In other words, we propose the following LO.

$$LO_{proposed}(t)=A_{LO}\cos\left(2\pi f_{IF\ beat\ leakage}t+\theta_{IF\ leakage}+\varphi_{LO}(t)\right). \quad (12)$$

If (5) is mixed with (9) and passes a low pass filter, then the final signal, $z(t)$, through the proposed SPC technique can be expressed as follows:

$$\begin{aligned}z(t)&=\underbrace{A_{leakage}\cos\left(\varphi_{leakage}(t)\right)}_{Leakage}\\&\quad\underbrace{+\sum_{k=1}^{K}A_{targets,k}\cos\left(2\pi f_{beat\ targets,k}t+\theta'_{targets,k}+\varphi_{targets,k}(t)\right),}_{Targets}\end{aligned} \quad (13)$$

where $\theta'_{targets,k}=\theta_{IF\ targets,k}-\theta_{IF\ leakage}$.

In (13), now that the leakage signal does not have any frequency and any constant phase, the term of the major cause of the noise floor rise in (11) is removed. The only term associated with the phase noise of the leakage is the term which envelopes the phase noise of the leakage in the cosine function in (13). This term can be approximated to $A_{leakage}$, if the same approximation in (11) is applied. Then the leakage is represented as just a dc value. Therefore, the proposed technique can resolve the problem of the phase noise of the leakage.

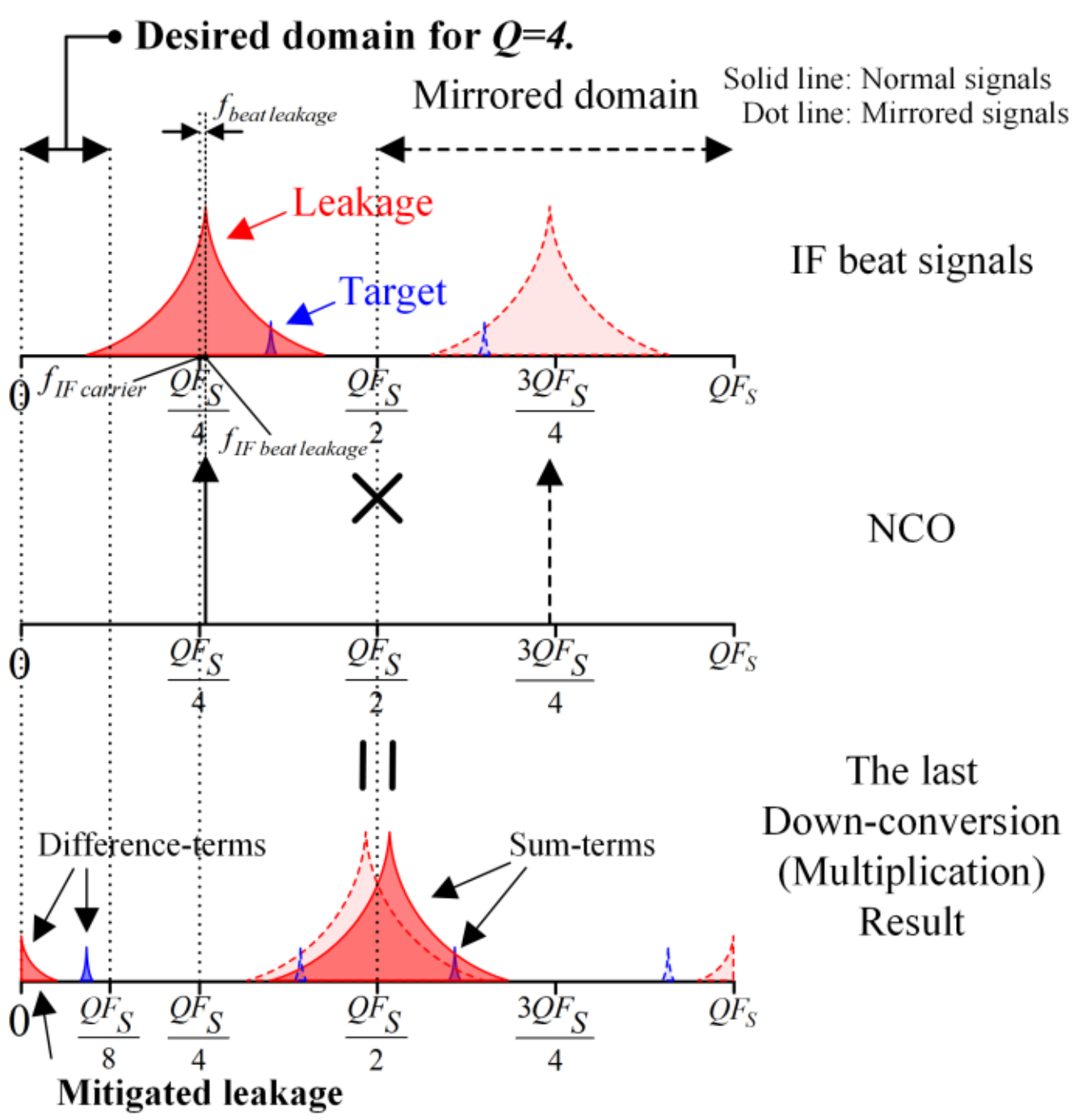

Fig. 3. Results of the final down-conversion through the SPC technique when the proposed frequency planning and oversampling is applied.

Even though we do not use the approximation, the effect of the proposed technique can be explained by Fig. 2. Through the SPC technique, the phase noise of the leakage is concentrated on the stationary point of the cosine function, because the leakage signal has no frequency and no constant phase. In case of the common method, the leakage signal, $y_{leakage}(t)$, has the beat frequency, thus the phase noise of the leakage can tremble at every point on the cosine function, which includes the maximum slope points. On the other hand, in case of the SPC technique, the phase noise of the leakage trembles only at the stationary point. Therefore, the magnitude of the phase noise is significantly decreased, and the mitigation of the leakage signal is possible.

Also, thanks to the SPC technique, the beat frequency of the leakage, $f_{beat\ leakage}$, due to the total internal delay is removed in the beat signals of the targets. Thus, the distance information of targets becomes more accurate. Although the constant phases of the targets' beat signals are changed due to the SPC technique, its influence on those power is insignificant.

## III. Implementation of Stationary Point Concentration Technique

### A. Strategic Frequency Planning and Oversampling

The SPC technique is implemented through the DSP. After generating a digital NCO whose frequency and constant phase are those of the IF beat signal of the leakage, the multiplication is carried out for the last down-conversion. However, the sampling and the frequency planning for the IF beat signals should be conducted carefully first. Once the digital bandwidth is decided, twice the bandwidth is going to be the minimum available sampling frequency by the Nyquist theorem. However, if the multiplication is carried out with this sampling frequency, the undesired sum-terms or mirrored of these are included in the desired frequency domain.

Therefore, strategic frequency planning and oversampling are applied in the SPC technique, and these are depicted in Fig. 3. The IF beat frequency, $f_{IF\ beat\ leakage}$, is placed around a quarter of the oversampled frequency domain. In this way, we can keep the desired domain as far away as possible from the sum-terms and the mirrored terms of these. Also, these undesired terms are located around the center of the digital IF frequency domain, which means that the sum-terms now can be certainly and easily removed by a digital low pass filter (LPF).

As the oversampling factor that is a positive rational number, $Q$, increases, the sum-terms and the mirrored terms of these are pushed further away from the desired domain, preventing any harmful influence. However, if $Q$ is too large, the cost of ADC increases for the high sampling rate. Small factor also gives a trouble. For example, if $Q$ is 2, it can encounter the aliasing of the sum-terms. Therefore, it is required to determine the appropriate oversampling factor by considering the engineer's own radar designs.

It is difficult to predict the $f_{beat\ leakage}$ before manufacturing an FMCW radar. However, $f_{IF\ beat\ leakage}$ is mostly occupied by $f_{IF\ carrier}$, because $f_{IF\ carrier}$ is usually much larger than $f_{beat\ leakage}$. Therefore, the frequency planning that locates $f_{IF\ carrier}$ instead of $f_{IF\ beat\ leakage}$ on a quarter of the oversampled frequency domain is reasonable. This explanation is reflected in Fig. 3. $f_{IF\ carrier}$ is $QF_S/4$, and the IF beat signal of the leakage whose frequency is $f_{IF\ beat\ leakage}$ slightly misses $QF_S/4$ point. Nevertheless, the sum-terms are around the center of the oversampled frequency domain, so the digital LPF can remove these obviously. Even in the special cases when long internal delay, large $f_{beat\ leakage}$, is inevitable [4], [8], the proposed technique also can be applied by increasing $Q$ a bit.

If we consider the undersampling which is also called as the bandpass sampling, the frequency planning can be realized on other locations, $QF_S\,(4N+1)/4$, where $N$ is a natural number.

### B. Final Down-conversion with Digital NCO

The block diagram of the SPC technique is shown in Fig. 4. The oversampled IF beat signal, $x[n]$, can be written as follows:

$$x[n]=\frac{A_S A_L}{2}\cos\left(2\pi f_{IF\ beat\ leakage}\,n\frac{T_S}{Q}+\theta_{IF\ leakage}\right.$$
$$\left.+\varphi_{IF\ leakage}(n\frac{T_S}{Q})\right)$$
$$+\sum_{k=1}^{K}\frac{A_S A_{T,k}}{2}\cos\left(2\pi\left(f_{IF\ beat\ leakage}+f_{beat\ targets,k}\right)n\frac{T_S}{Q}\right.$$

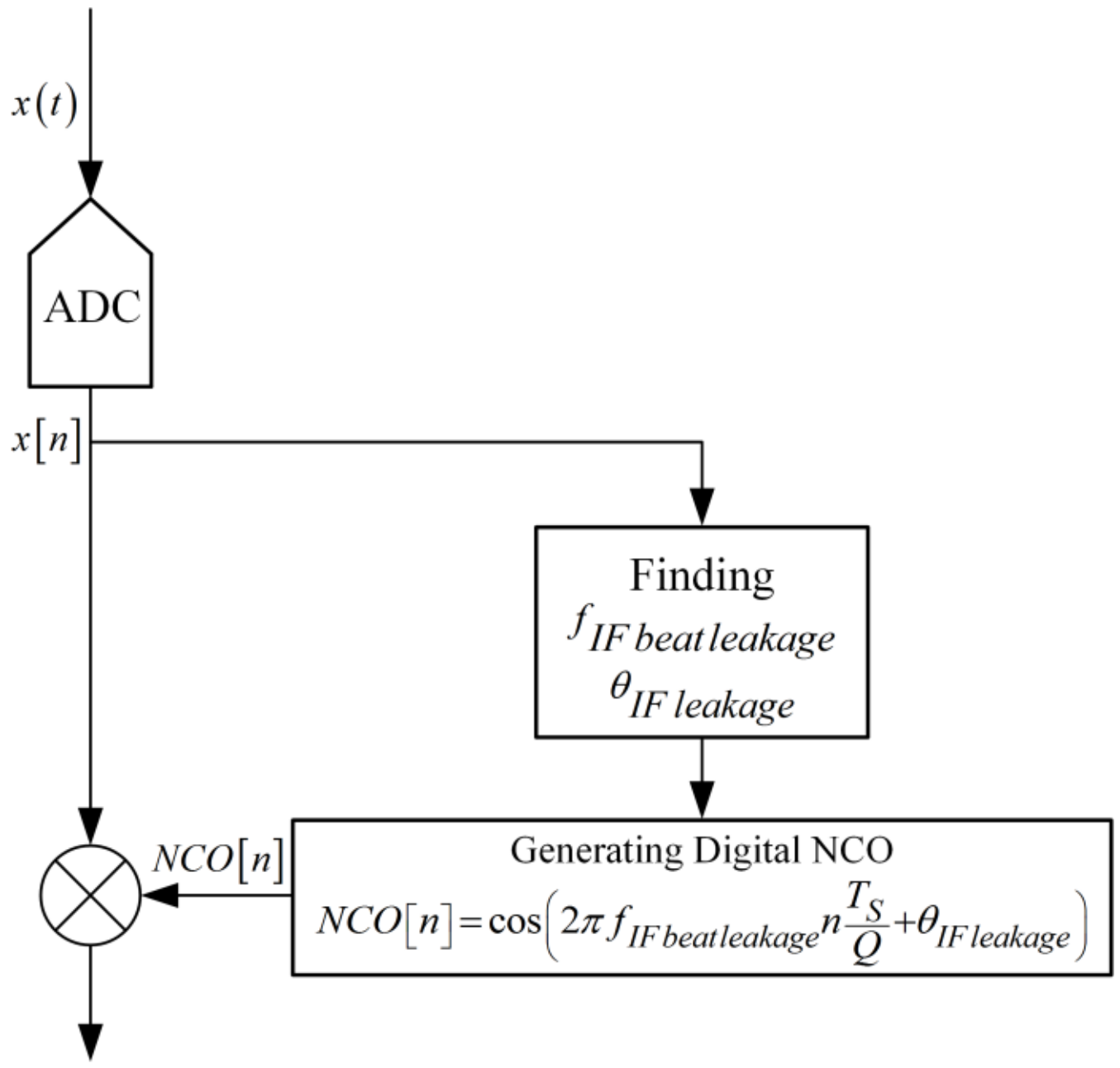

Fig. 4. Block diagram of the SPC technique.

$$+\theta_{IF\ targets,k} + \varphi_{IF\ targets,k}(n\frac{T_S}{Q})\Big), \quad (14)$$

where $T_S/Q = 1/QF_S$ is the oversampling interval. The excessively large power of the leakage signal is an obvious problem. However, we change this problem to a solution for finding $f_{IF\ beat\ leakage}$ and $\theta_{IF\ leakage}$. We use the fact that the magnitude of the leakage signal is much larger than that of the target signal [12], [15]. After applying the fast Fourier transform (FFT) to $x[n]$, we transform the FFT result to the power spectrum and the phase response. When the FFT is applied, we use the zero-padding to minimize the error caused by insufficient spacing in the frequency domain. The form of the FFT with the zero-padding converges to that of the discrete time Fourier transform (DTFT). This proves that the zero-padding helps find the real location of the signal peak [23], [24]. Using this advantage of the zero-padding, we find the peak which represents the maximum power around a quarter of the oversampled frequency domain. The index number, $k_{IF\ leakage}$, from the maximum peak can be extracted. Finally, we can find out $f_{IF\ beat\ leakage}$ and $\theta_{IF\ leakage}$ as follows:

$$\begin{aligned} k_{IF\ leakage} &= \underset{\frac{QF_S}{4}<k<\frac{QF_S}{2}}{\arg\max} \left|X[k]\right|^2, \\ f_{IF\ beat\ leakage} &= \frac{QF_S}{NFFT}\left[k_{IF\ leakage} - 1\right], \\ \theta_{IF\ leakage} &= \angle X\left[k_{IF\ leakage}\right], \end{aligned} \quad (15)$$

where $X[k]$ is the result of $NFFT$-point FFT of $x[n]$, $NFFT$ is the total number of real samples and zero-pads, $QF_S$ is the sampling frequency for the oversampling, and $\angle X$ is the phase response of $X[k]$.

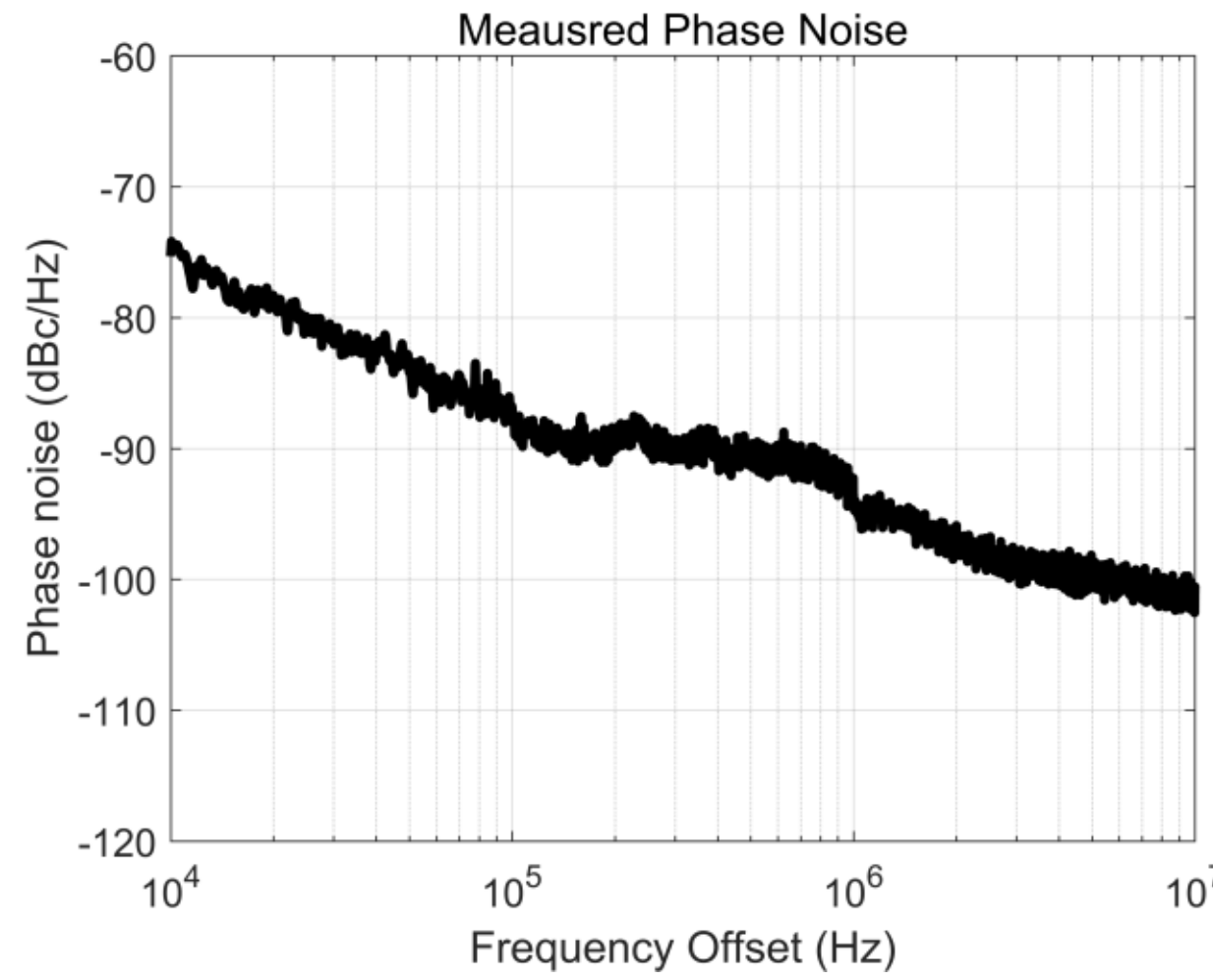

Fig. 5. Measured PSD of the phase noise. The result is the average of 10 measurements taken with the spectrum analyzer, Agilent Technologies E4440A.

With the extracted $f_{IF\ beat\ leakage}$ and $\theta_{IF\ leakage}$, the following digital NCO can be generated.

$$NCO[n] = \cos\left(2\pi f_{IF\ beat\ leakage} n\frac{T_S}{Q} + \theta_{IF\ leakage}\right) \quad (16)$$

Then, the last down-conversion can be conducted by multiplying $x[n]$ and $NCO[n]$. After the multiplication, the desired terms can be expressed as follows:

$$\begin{aligned} \psi[n] = & \underbrace{\frac{A_S A_L}{4}\cos\left(\varphi_{IF\ leakage}(n\frac{T_S}{Q})\right)}_{Leakage} \\ & \underbrace{+\frac{A_S A_{T,k}}{4}\sum_{k=1}^{K}\cos\Big(2\pi f_{beat\ targets,k}\, n\frac{T_S}{Q} + \theta'_{targets,k} + \varphi_{IF\ targets,k}(n\frac{T_S}{Q})\Big)}_{Targets}. \end{aligned} \quad (17)$$

Then, a digital LPF and the decimation can be added depending on the design.

## IV. Simulation

In this Section, the simulations are carried out for the analysis of the SPC technique and the prediction of the experiment results. An actual PSD of the phase noise from the measurement is usually used for a better reliability of the simulation [16], [25]. Also, FMCW radar simulation was done in IF domain for the fast simulation in [25]. Referring to these, in this paper, the PSD of the phase noise which is measured in the RX IF stage is used for the reliable simulation. Based on this phase noise information which can be seen in Fig. 5, all simulations were conducted in digital IF domain.

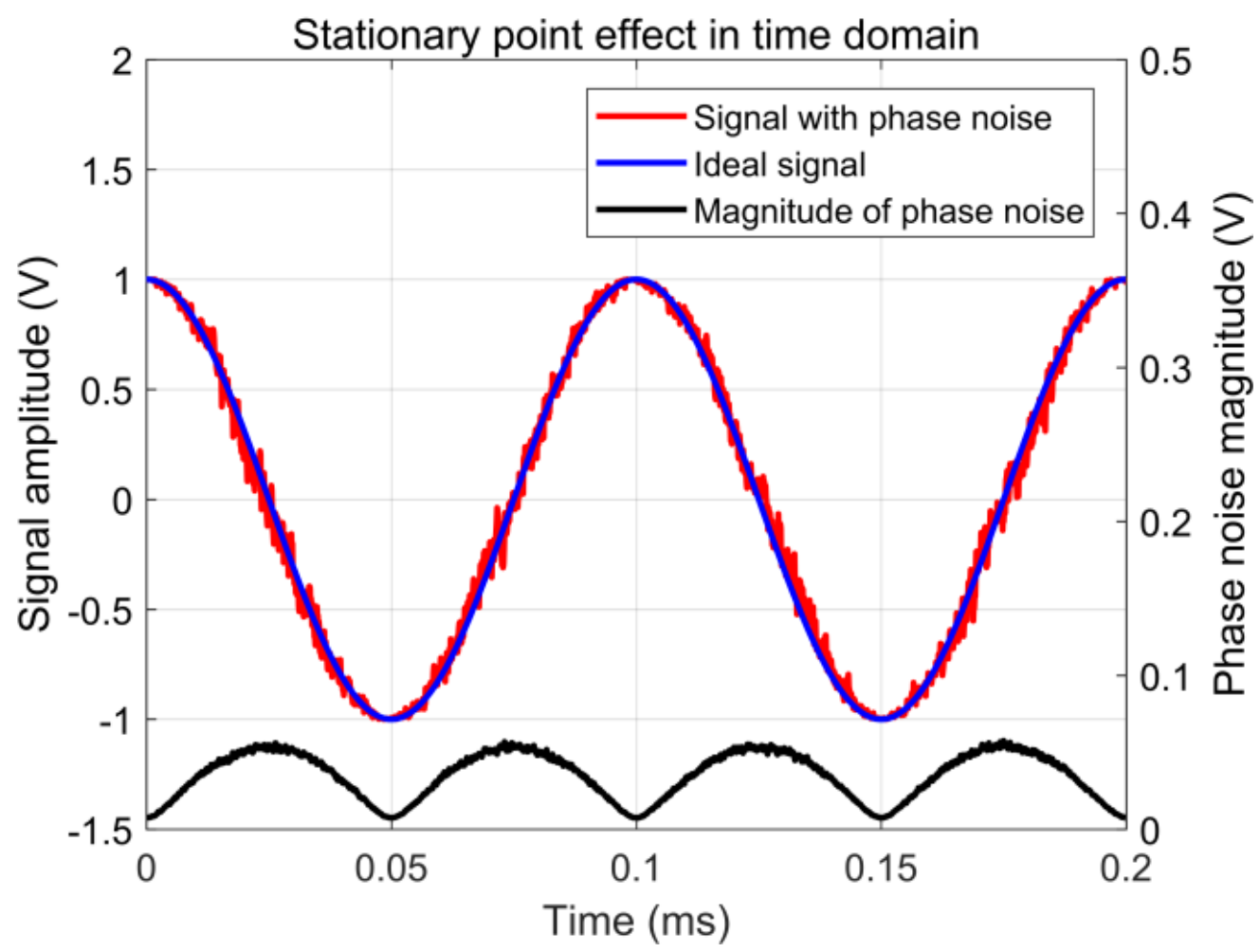

Fig. 6. Stationary point effect of phase noise in time domain. The results are the average of 1000 simulations.

TABLE I
PARAMETERS OF HETERODYNE FMCW RADAR FOR SIMULATION/EXPERIMENT

| Parameters | Values |
|---|---|
| Sweep period ($T$) | 880 us |
| Desired digital bandwidth | 1.25 MHz |
| Minimum available sampling frequency ($F_S$) | 2.5 MHz |
| Oversampling factor ($Q$) | 4 |
| Oversampling frequency ($QF_S$) | 10 MHz |
| # of samples in a chirp | 8800 |
| # of samples after discarding early part in a chirp | 8192 |
| IF carrier frequency ($f_{IF\ carrier}$) | 2.5 MHz |
| Window | Hann |
| $NFFT$ for finding $f_{IF\ beat\ leakage}$ and $\theta_{IF\ leakage}$ | $2^{20}$ |
| Desired maximum detectable range | 1100 m |
| Apparent range resolution | 1.074 m |

## *A. Stationary Point Effect Analysis in Time domain*

The explanation about the stationary points and the maximum slope points in Fig. 2 and Section II-*B* can be verified by the simulations in the time domain. A simple example is shown in Fig. 6. For this example, two sinusoidal signals whose frequencies are 10 kHz and amplitudes are 1 were created with the sampling rate of 10 MHz. One includes the phase noise, and the other is just an ideal signal. The phase noise which is manifested as the voltage noise can be extracted by subtracting the ideal signal from the signal with the phase noise. We added the magnitude of the extracted phase noise on the graph. As shown in Fig. 6, the magnitude of the phase noise approaches the local maximum as the phase goes to the maximum slope points, and it becomes the local minimum as the phase goes to the stationary points. The SPC technique makes good use of these properties. By concentrating the phase noise of the leakage on the stationary point, the SPC technique can significantly mitigate the magnitude of it.

## *B. Performance Analysis of SPC Technique*

Before the simulation of the small drone detection, we simulate only the leakage signal and check the degree it is mitigated by the SPC technique in the power spectrum. If the theory is properly implemented, the magnitude of the leakage's phase noise should decrease and so will the noise floor. The parameters of the heterodyne FMCW radar are listed in Table I. In order to compare the experiment results with the simulation results, the same parameters were used in the simulations. In considering the value of $NFFT$, larger $NFFT$ would accordingly yield more accurate $f_{IF\ beat\ leakage}$ and $\theta_{IF\ leakage}$. However, excessive $NFFT$ causes an overload to the computation process. In this paper, $NFFT$ of $2^{20}$ is used. This gives the FFT resolution of 9.54 Hz based on the oversampling frequency. Thus, the maximum error of $f_{IF\ beat\ leakage}$ due to the FFT resolution is about 4.77 Hz. In this work, $f_{IF\ carrier}$ is 2.5 MHz. If $f_{beat\ leakage}$ is 10 kHz, $f_{IF\ beat\ leakage}$ is 2.51 MHz. Thus, the error ratio of $f_{IF\ beat\ leakage}$ due to the FFT resolution is only about 0.00019 %. Therefore, the zero-padding we conducted provides sufficient FFT resolution to find accurate $f_{IF\ beat\ leakage}$ and $\theta_{IF\ leakage}$. In terms of the computation time, an Intel N3060 CPU and a 4 GB RAM yields only about 0.095 s.

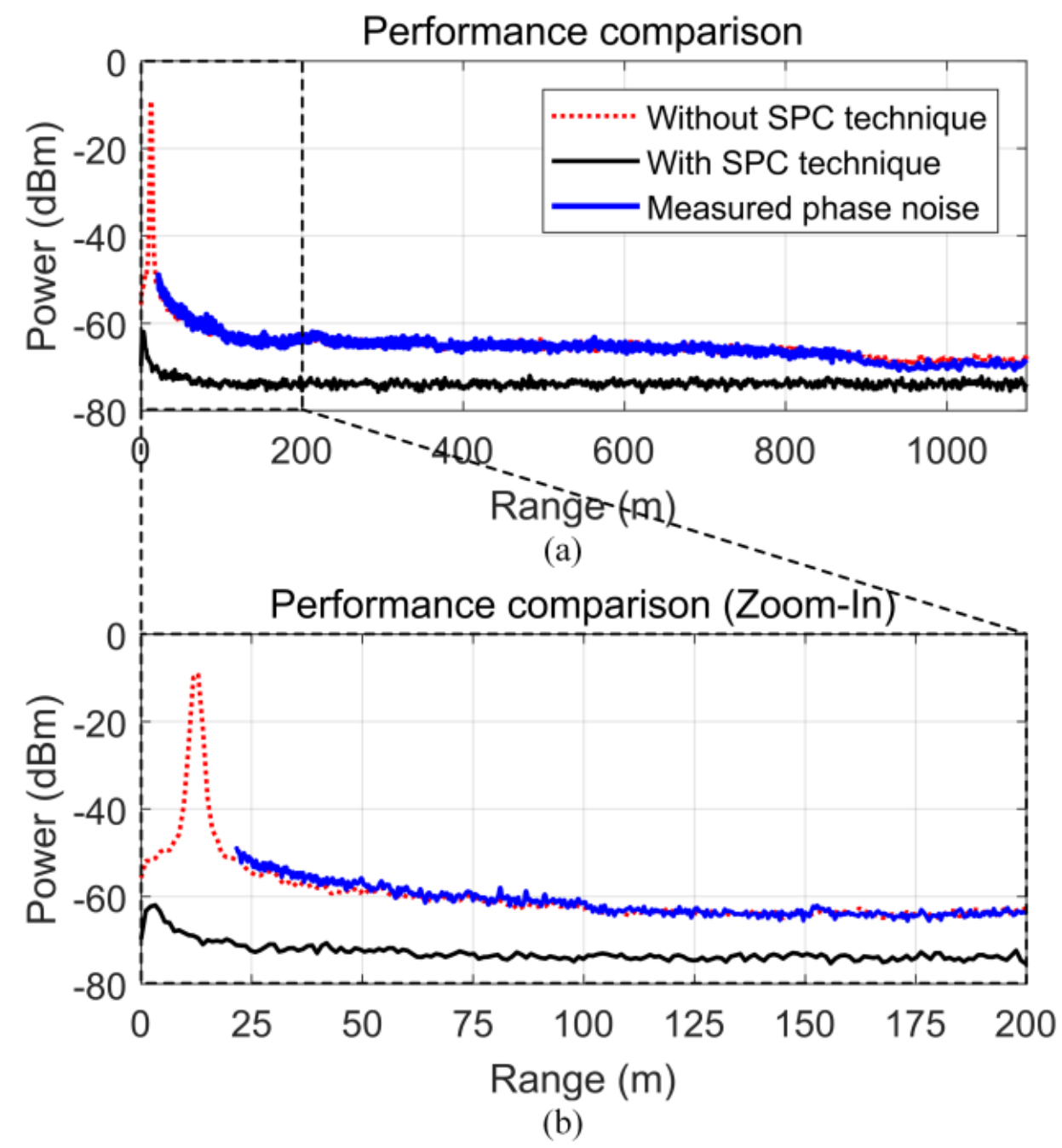

Fig. 7. Simulation results of leakage mitigation. (a) shows the performance comparison of the power spectrum. (b) is the zoom-in version of (a). The results are the average of 100 power spectra.

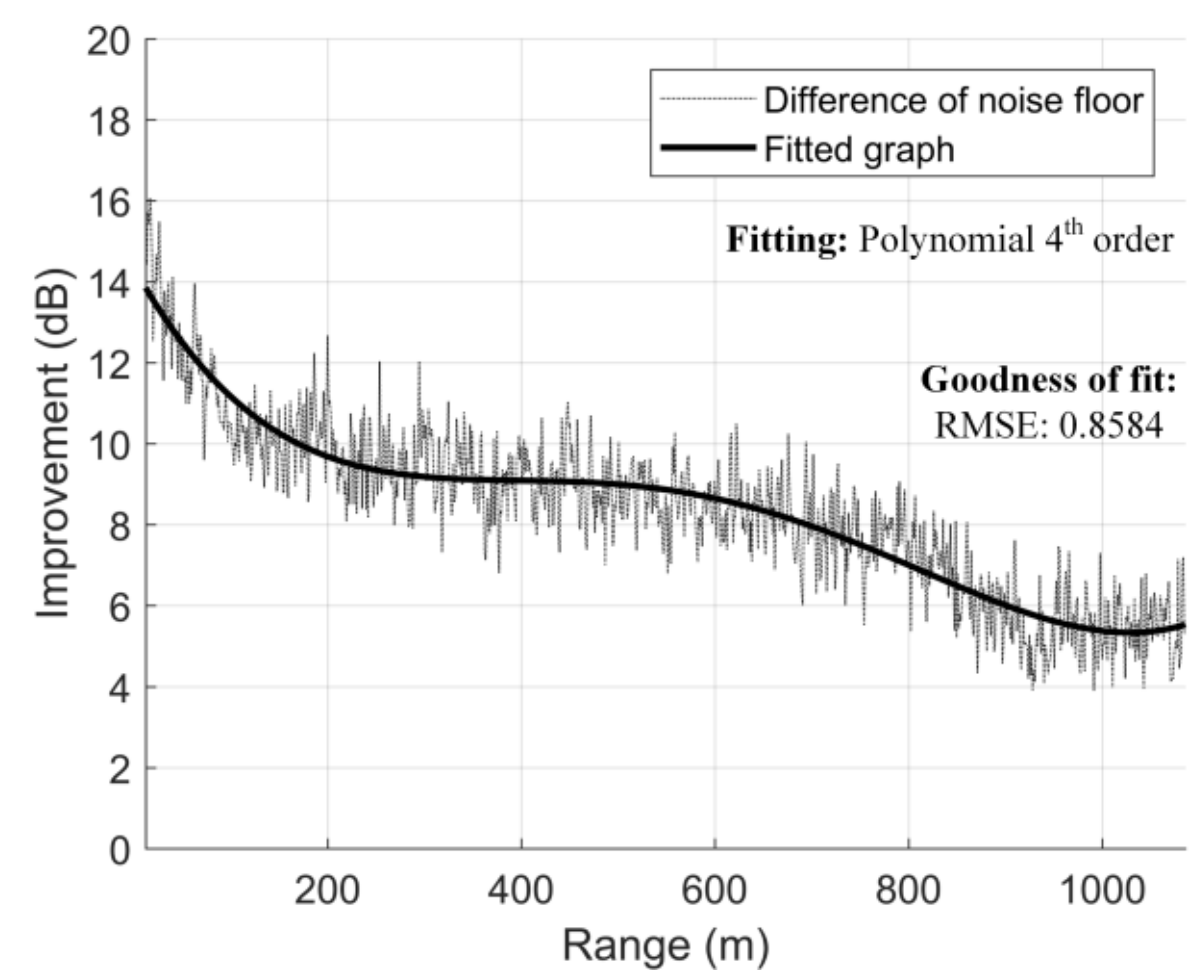

Fig. 8. Improvement performance of SPC technique in simulation.

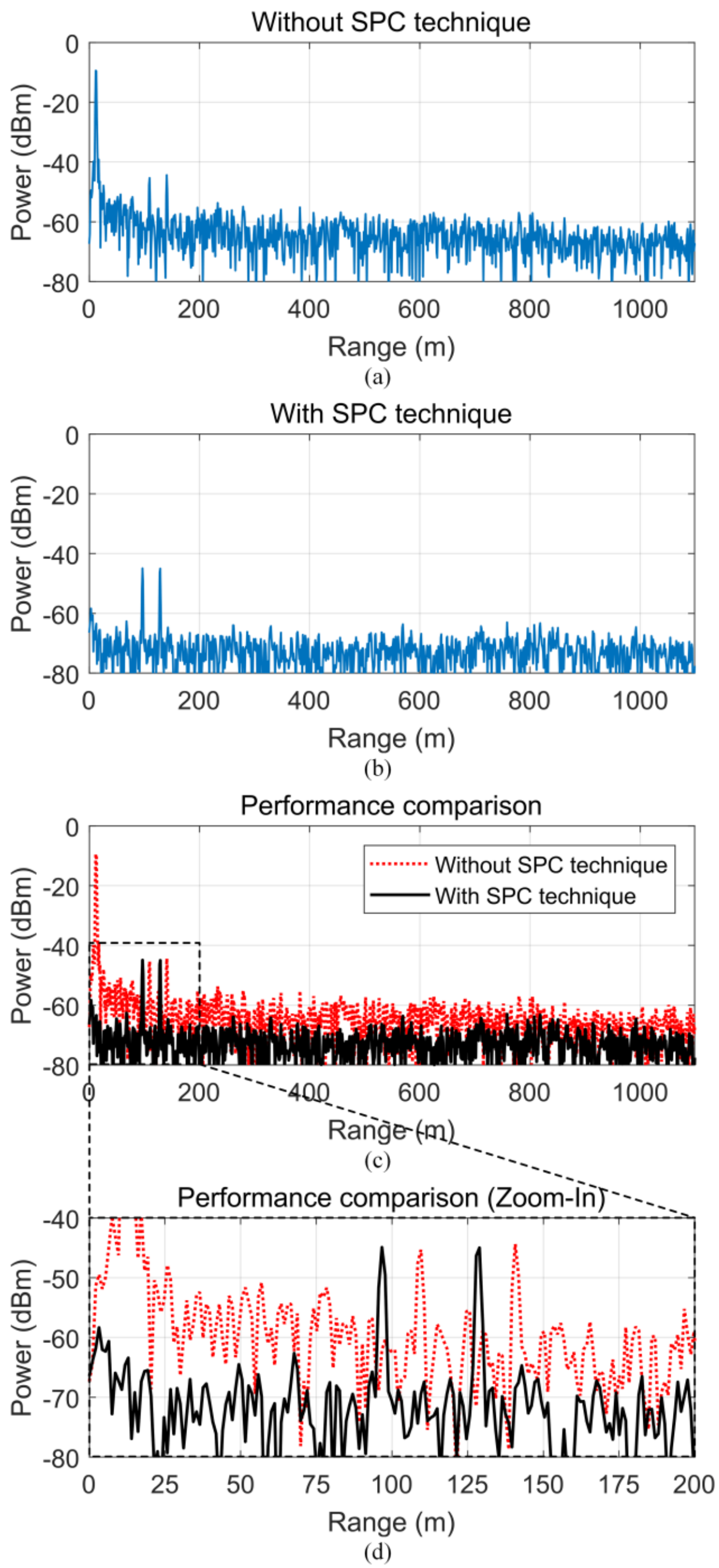

Fig. 9. Simulation results of target detection with leakage mitigation. (a) and (b) are the power spectra from the common method without the SPC technique and from the SPC technique, respectively. (c) shows the comparison of the power spectrum. (d) is the zoom-in version of (c).

Fig. 7 shows the performance of the SPC technique in the power spectrum. For a clear comparison, we took an average on the power spectra from 100 chirps, which reduced the variance of the noise floor. The entire power spectrum is shifted to the left by the beat frequency of the leakage because the SPC technique involves a subtraction of the beat frequency. The remarkable mitigation of the leakage is apparent in Fig .7. As can be observed in the figure, the measured phase noise is in concordance with the noise floor. Thus, the simulation reflects the measured phase noise accurately. From the results, we can expect that the proposed SPC technique will mitigate the high noise floor caused by the phase noise of the leakage. The improvement performance of the SPC technique is displayed more clearly in Fig. 8. In the figure, the graph displays the difference of noise floor whose subtraction was done after the alignment of the two power spectra. In order to estimate more specific value of the improvement, a fitted curve is overlaid. The maximum and the minimum degrees of the improvement are estimated to be around 13.8 dB and 5.3 dB, respectively.

### *C. Target Detection with SPC Technique*

The simulations of the target detection were conducted considering the experiment plans in Section VI. We consider the real-time target detection, which is the practical application. Thus, we show the results using only one chirp without the averaging or the zero-padding. In the simulations, two target signals were added. An adequate signal power was set to check the clear improvement in the SNR. If properly implemented, the SNR of the target signals will increase as the noise floor decreases. The expected result is depicted in Fig. 9. As can be seen from the figure, the SNR of the target signals are significantly increased by the SPC technique. Because of the same reason that we mentioned in the previous subsection, the power spectrum from the SPC technique is shifted to the left. Also, as we explained in Section II-*B*, there is little influence on the power of the target signals by the changed constant phases from the SPC technique.

## V. Radar System

In verifying the performance of the proposed SPC technique through the experiments, a *Ku*-band heterodyne FMCW radar was used. The *Ku*-band heterodyne FMCW radar has been continuously upgraded for drone detection by Microwave & Antenna laboratory in Korea Advanced Institute of Science and Technology (KAIST). The specifications of the radar are listed in Table II.

Fig. 10 shows the block diagram of the radar. The radar mainly consists of baseband (BB) & IF stages, block up converter (BUC), low-noise block (LNB), and antennas. The BB and IF stages are placed together in a metal case. Analog Device AD9854, a direct digital synthesizer (DDS), is used to generate the LFM signal with high linearity. For the TX and RX antennas, corrugated conical horn antennas were used. Corrugated conical horn antenna provides high-efficiency and rotationally symmetric beam pattern [26]. The radar includes a software defined radio (SDR), Ettus USRP N210. In the experiments, we used USRP N210 for the oversampling and the positioning of $f_{IF\ carrier}$. The remaining procedures of the proposed technique after the oversampling were conducted through MATLAB in a mini-PC.

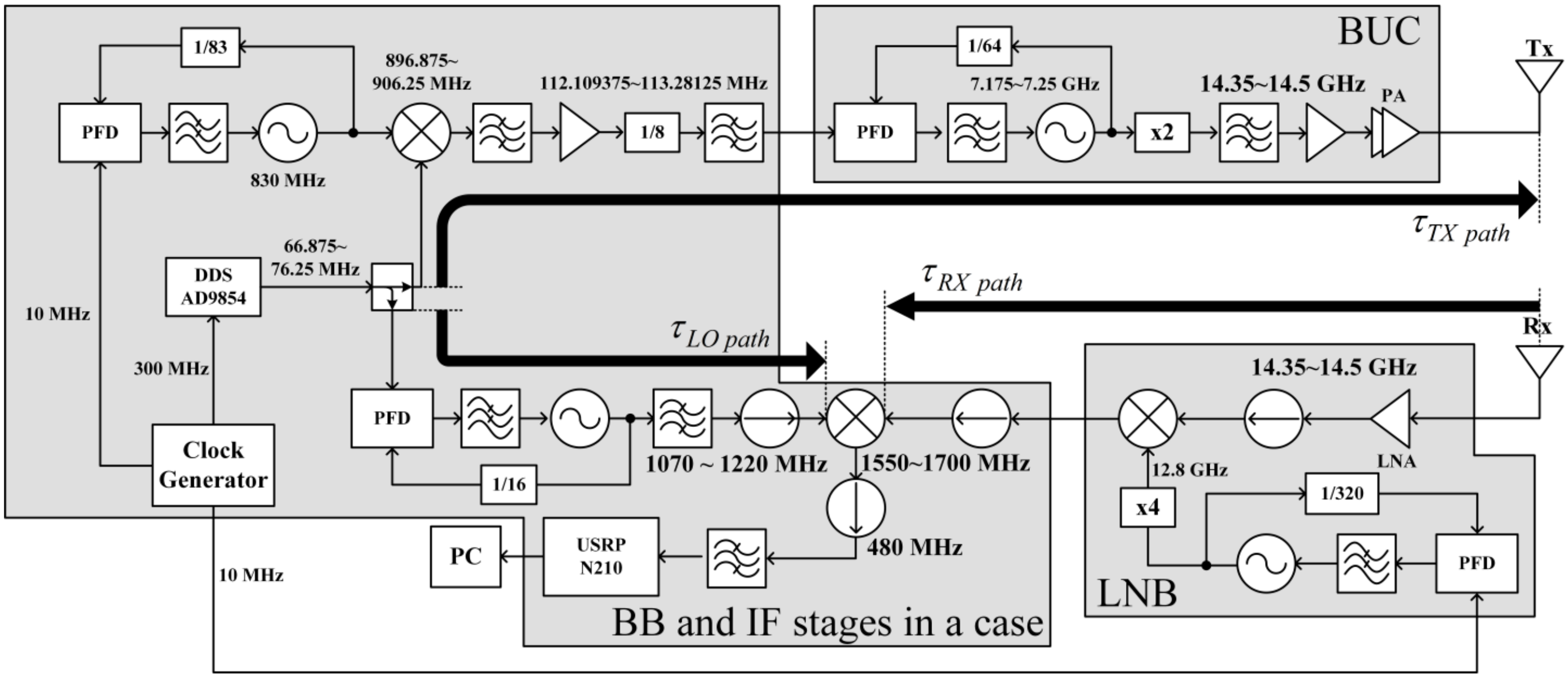

Fig. 10. Block diagram of the *Ku*-band heterodyne FMCW radar system.

TABLE II
SPECIFICATIONS OF THE *KU*-BAND HETERODYNE FMCW RADAR

| Property | Value | Units |
|---|---|---|
| Operating frequency | 14.35-14.50 | GHz |
| Transmit power | 30 | dBm |
| Antenna gain | 16 | dBi |
| Sweep bandwidth | 150 | MHz |
| Range resolution | 1 | m |

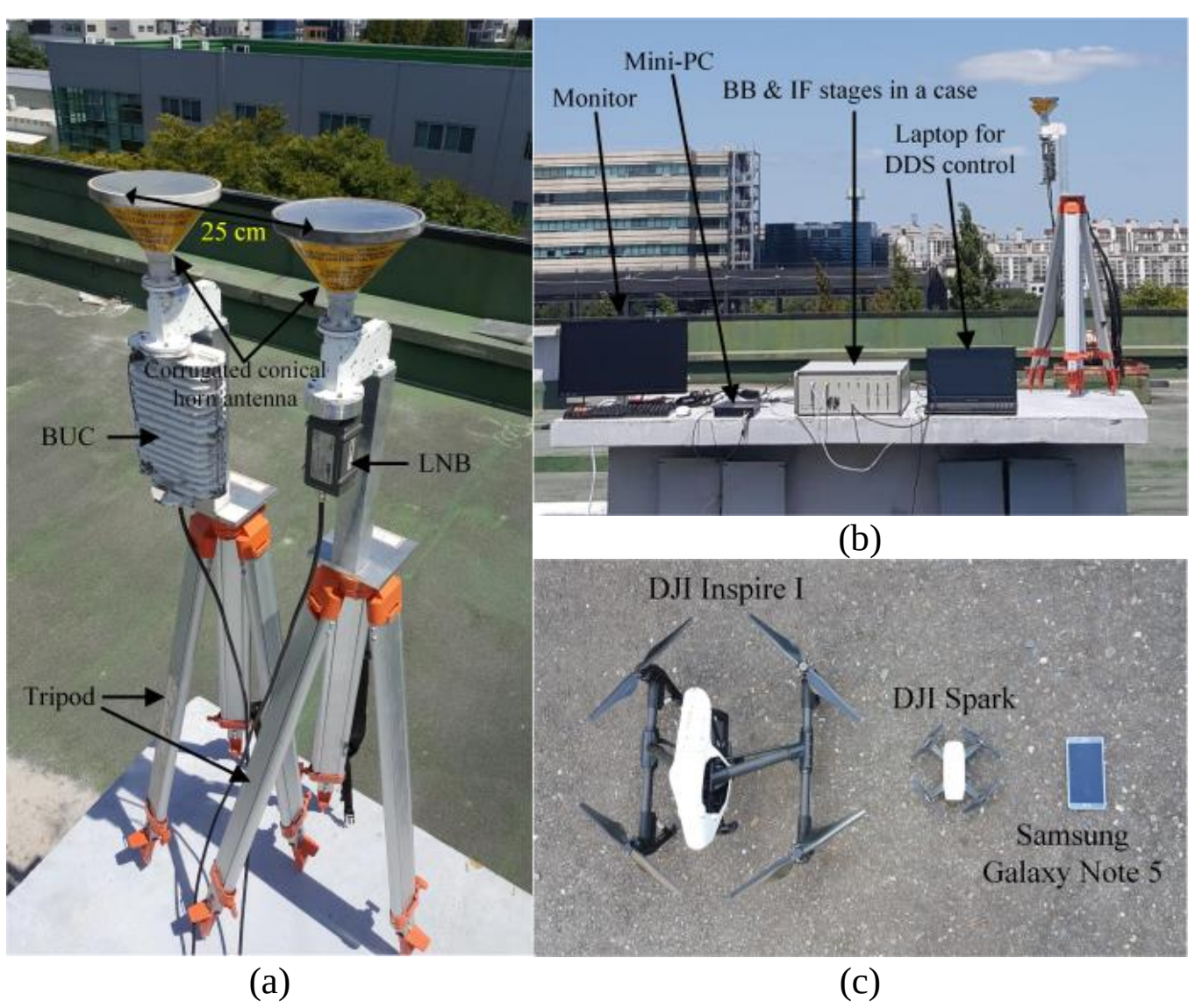

Fig. 11. Radar set-up and targets. (a) shows RF front-end and (b) shows the other sections of the radar. The small drones, DJI Spark and DJI Inspire I are shown in (c).

## VI. EXPERIMENTS

Radar set-up for the experiments is shown in Fig. 11. The quasi-monostatic configuration was chosen. The distance between two antennas was 25 cm. The antennas were set to look up at the sky and the system was installed at the rooftop of a building in KAIST. Therefore, we created a situation that the received signals contain only the leakage and the reflected wave by the targets that we set. For the targets, we used the small drones, DJI Spark and DJI Inspire I. We performed two experiments, *Experiment A* and *Experiment B*, to demonstrate the proposed SPC technique.

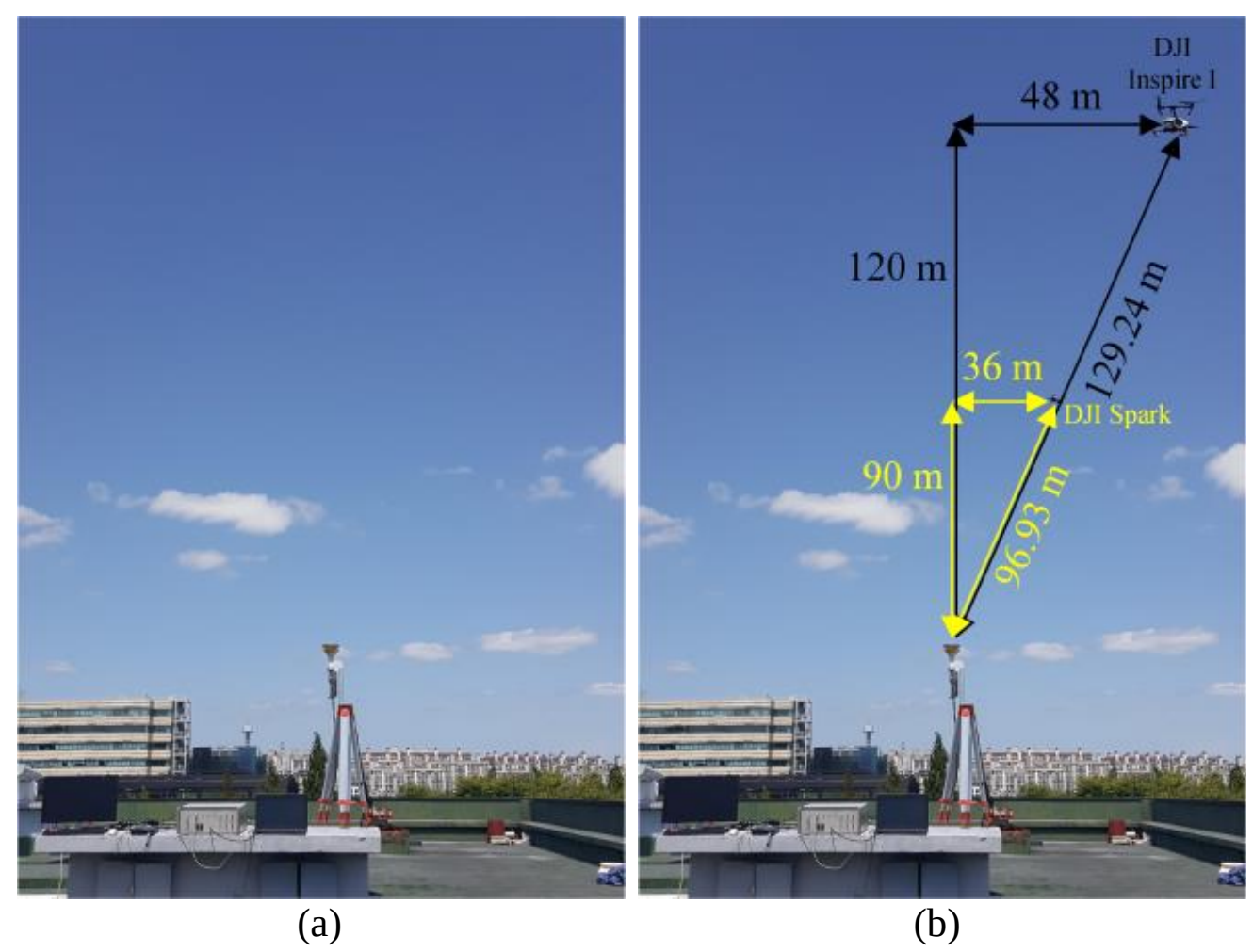

Fig. 12. Experiment scenarios. (a) *Experiment A,* (b) *Experiment B*.

In *Experiment A*, we focused on observing the reduced noise floor, which is directly related to the phase noise of the leakage signal. In order to achieve this purpose, we measured the signal in a situation with no target introduced to the radar, receiving only the leakage signal. By doing so, we will clearly see the comparison of the noise floors from the SPC technique and the common method. This experiment demonstrates the simulation analysis in Section IV-*B*. Fig 12-(a) displays the arrangement of the experiment.

In *Experiment B*, we intentionally placed the small drones at a slightly askew direction from the boresight. In this way, we can show the performance that even if the small drones try to invade through relatively weak beam area and the received signals are weak, the SPC technique can detect the small drones effectively. For the precise positioning of the drones, we used GPS information. This experiment demonstrates the simulation analysis in Section IV-*C*. Fig. 12-(b) shows the configuration of *Experiment B*.

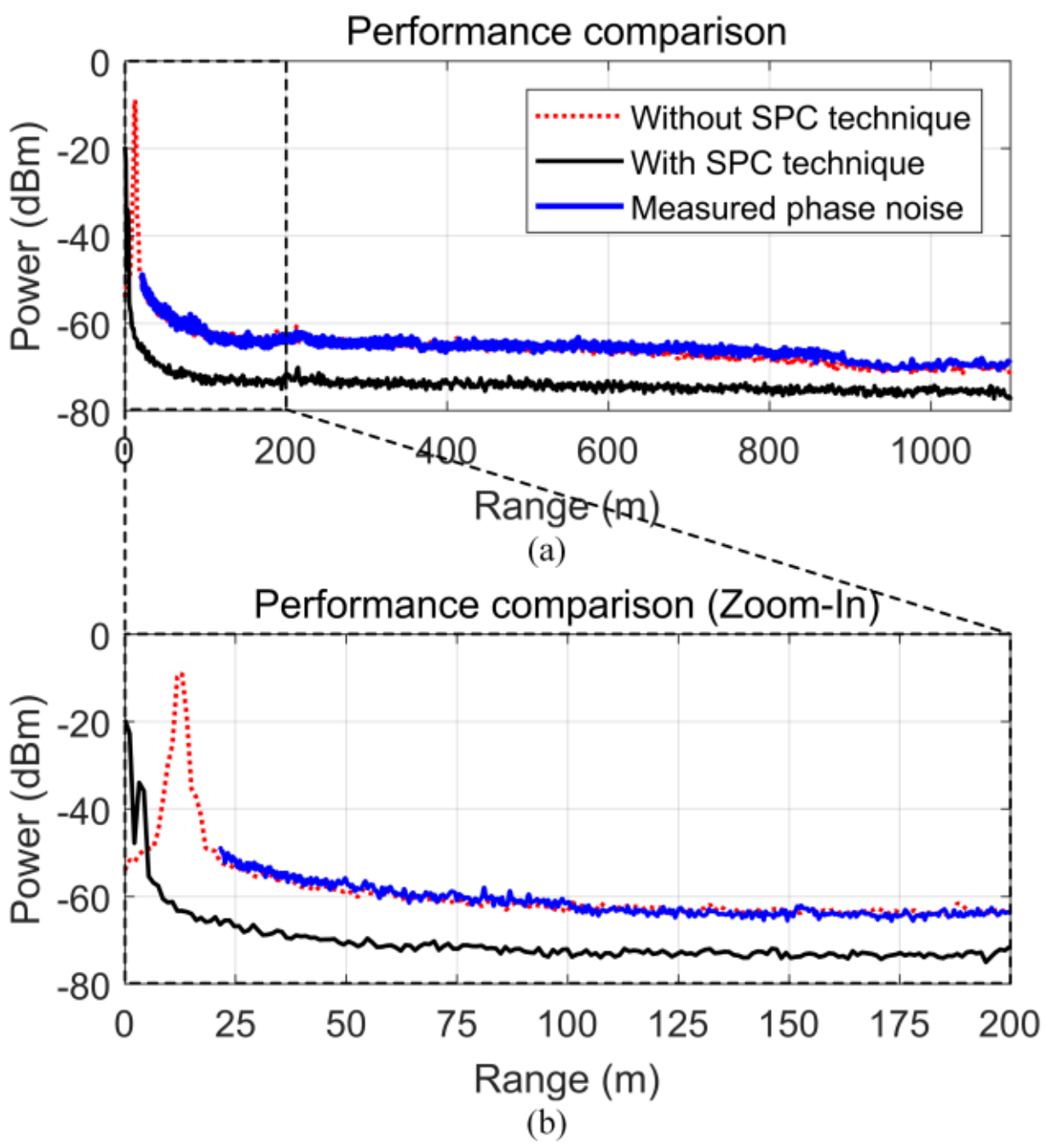

Fig. 13. Results of *Experiment A*. (a) shows the performance comparison of the power spectrum. (b) is the zoom-in version of (a). The results are the average of 100 power spectra.

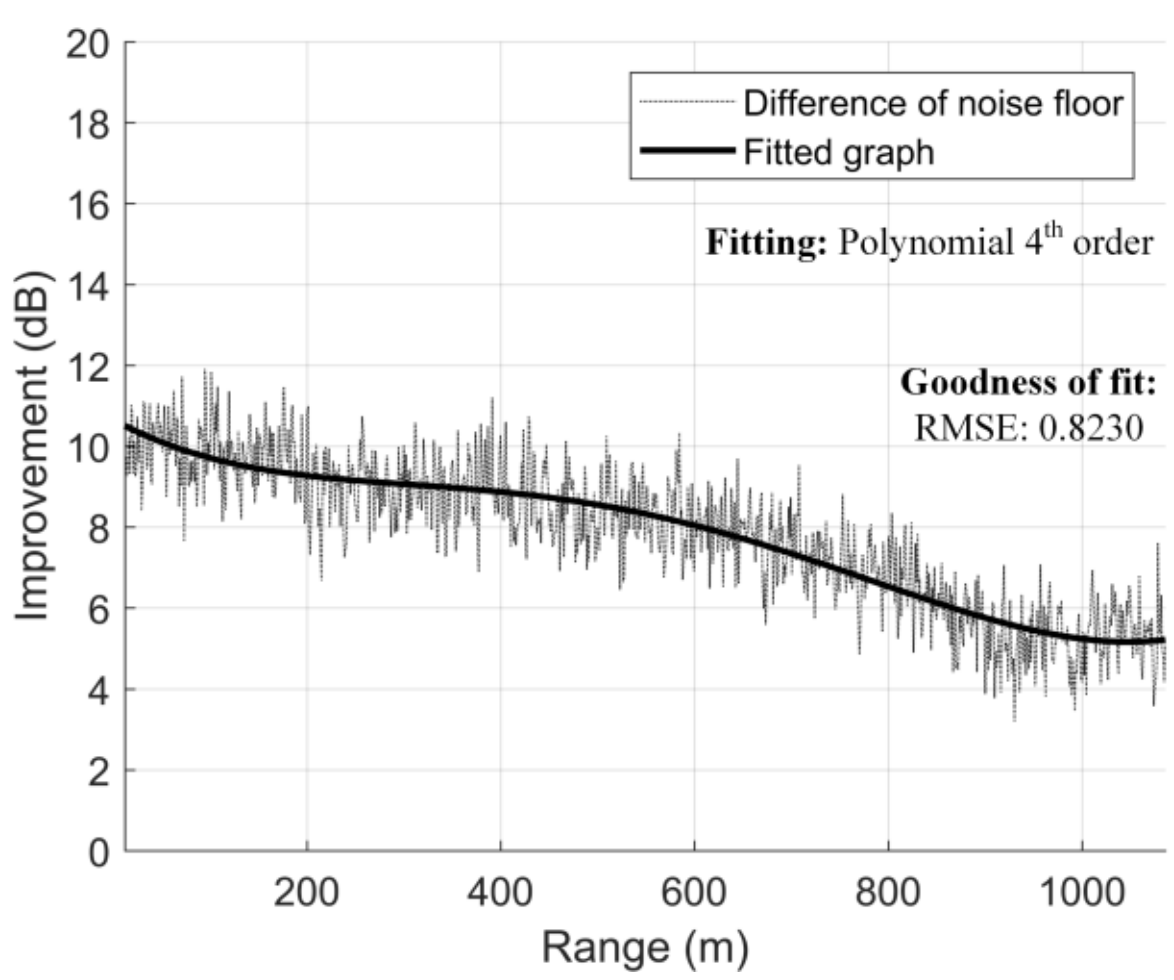

Fig. 14. Improvement performance of SPC technique in *Experiment A*.

## VII. Results and Discussion

### *A. Results and Discussion on Experiment A*

The results of *Experiment A* are shown in Fig. 13 and Fig. 14. As we did in the simulations, we took an average on the power spectra from 100 chirps for the clear comparison. In Fig. 13, the noise floor is well matched with the measured phase noise, and there is a noticeable improvement in the mitigation of the leakage. Therefore, it is verified that the proposed SPC technique reduces the noise floor caused by the phase noise of the leakage. In Fig. 14, the difference of noise floor graph is the subtraction of the two resulting power spectra in Fig. 13-(a). When the subtraction was done, the two power spectra were aligned. Also, the leakage peak is not considered in the improvement graph. The maximum and the minimum degrees of the improvement are measured to be about 10.5 dB and 5.2 dB, respectively.

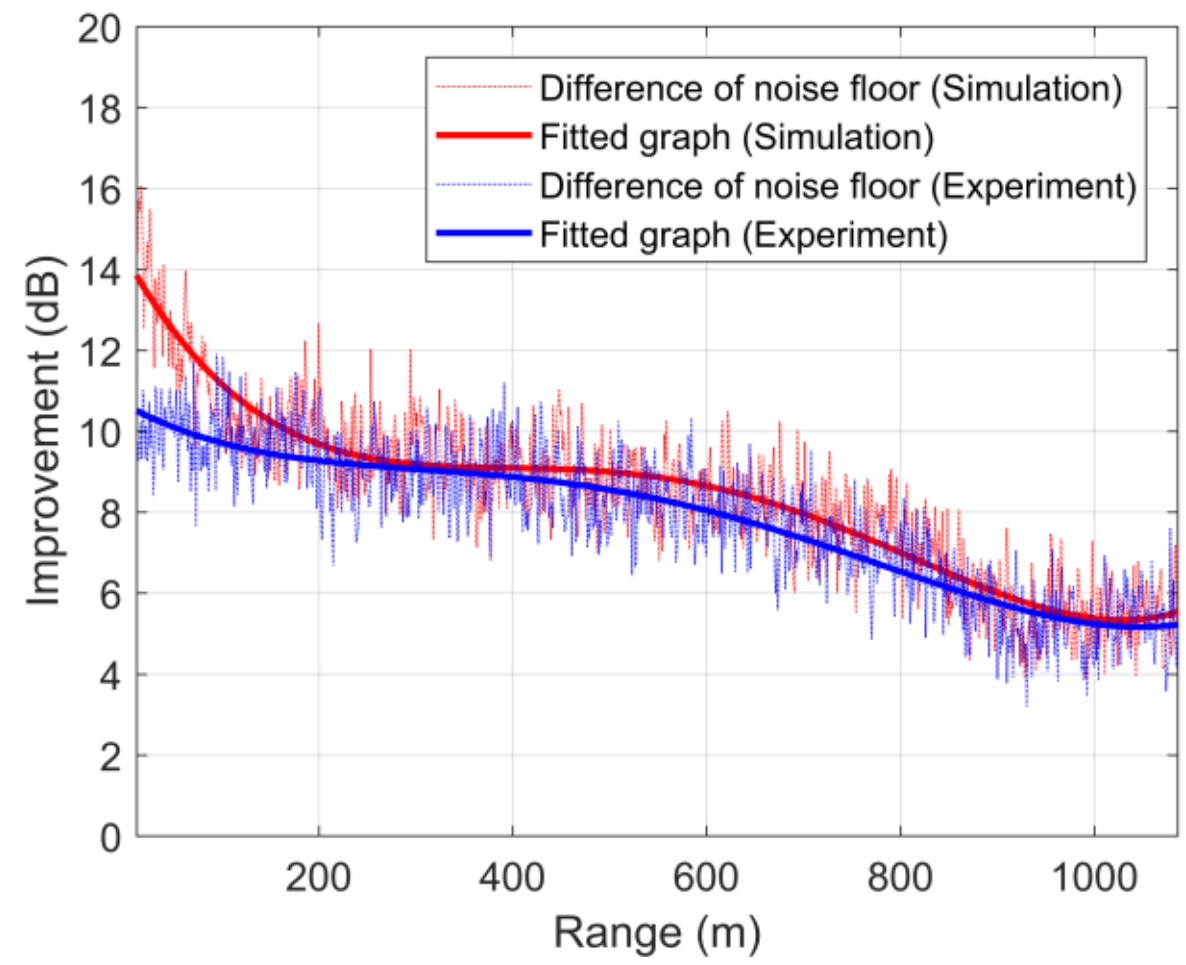

Fig. 15. Comparison between simulation and experiment results of improvement performance.

For the clear comparison with the simulations, we overlaid the improvement performance graphs in Fig. 15. The difference of the improvement between the simulation and the experiment is within 1dB in most range domain. However, in the near distance range, the difference is widened up to 3.3 dB. The reason of this can be explained by comparing the simulation result, Fig. 7-(b) and the experiment result, Fig. 13-(b). While the result in Fig. 7-(b) presents a complete leakage mitigation from a single dominant leakage assumption, in Fig. 13-(b), some components near the dc are observed. Also, the powers of the components, are less than the power of the leakage in the power spectrum without the SPC technique. Therefore, the components near dc in the experiment result, Fig. 13-(b), are not the dominant leakage but other minor leakages from minor leakage paths. Because of these other minor leakages, the expected improvement results from the simulations somewhat miss that of the experiment in the near-distance range. However, although the minor leakages may exist, the proposed SPC technique certainly mitigates the dominant leakage as we aimed in this paper, and this fact leads to the significantly improved power spectrum.

### *B. Results and Discussion on Experiment B*

Fig. 16 shows the results of *Experiment B*. Only one chirp was used for the results to perform the real-time target detection which is the practical application. Thus, the averaging and the zero-padding were not applied. As shown in Fig. 16, the SNR of the target signals are significantly increased. Based on these results, we can expect that the maximum detectable distance for the small drones will increase by the SPC technique. In a real situation, when we detect the drones that are moving away from the radar, the drones will first disappear from the common method based radar, before they disappear from the SPC technique based radar. In other words, when we try to detect the invading drones for the terrorism, the SPC technique based radar will detect the drones earlier than the common method based radar.

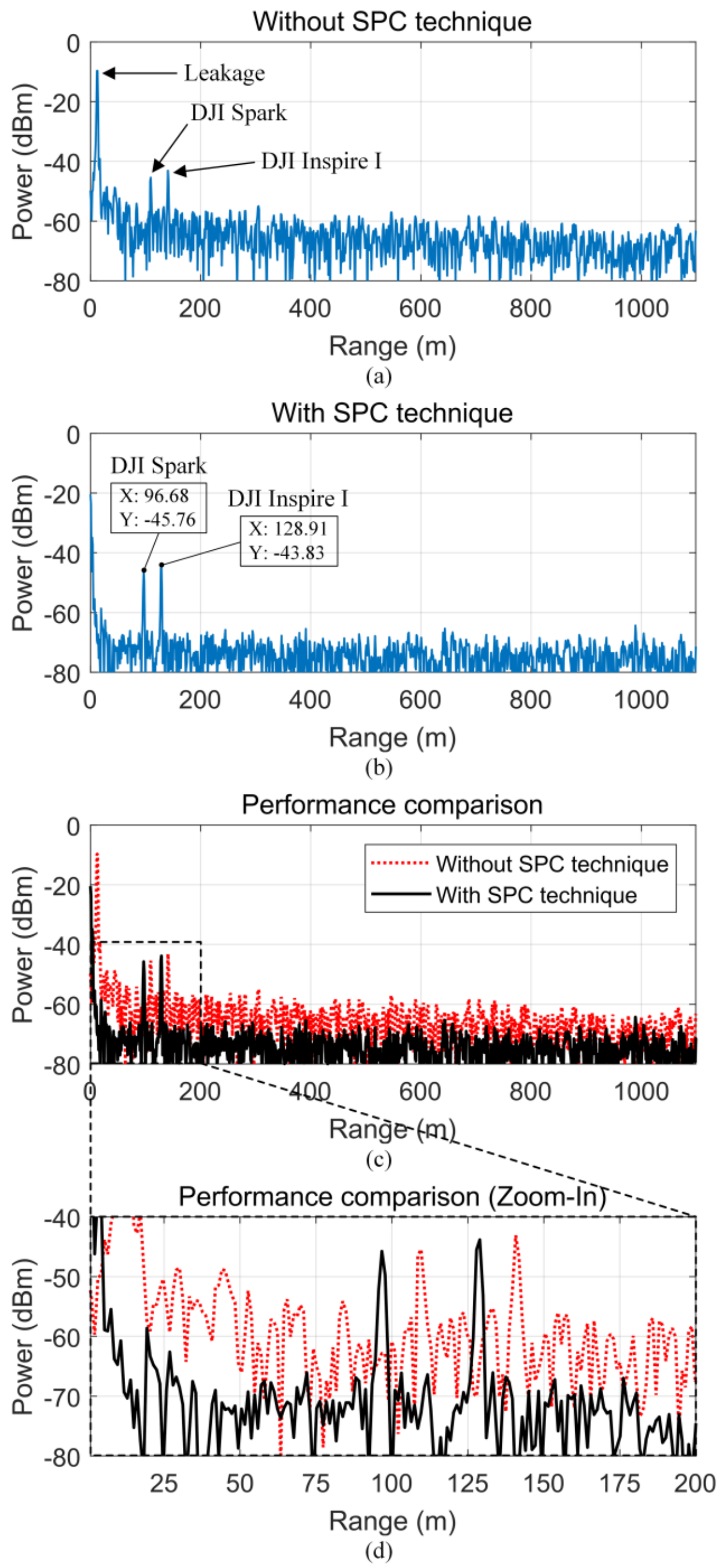

Fig. 16. Results of *Experiment B*. (a) and (b) are the power spectra from the common method without the SPC technique and from the SPC technique, respectively. (c) shows the comparison of the power spectrum. (d) is the zoom-in version of (c).

The measured distance information in Fig. 16-(b) is quite accurate when it is compared with the true distance in Fig. 12-(b). In the common method without the SPC technique, the received signals from the targets not only have the delay due to the target distance but also the total internal delay in the radar. This total internal delay corresponds to the beat frequency of the dominant leakage. On the other hand, in the SPC technique, because the beat frequency of the dominant leakage is deleted in the beat signals of the targets, only the beat frequencies of the targets remain in the beat signals of the targets. This effect of the SPC technique results in the shift of the power spectrum to the left and provides more precise distance information than the common method. In Fig. 16-(d), it is shown that the powers of the targets remain the same. This verifies the theory and the simulation results that there is little influence though the constant phases in the beat signals of the targets are changed with the SPC technique.

## VIII. Conclusion

The SPC technique to mitigate the dominant leakage in the heterodyne FMCW radar for the small drone detection has been explained and demonstrated in detail with the theories, the simulations, and the experiments. Unlike the major trend in this research field, the proposed technique suggests a new approach to mitigate the leakage problem. Also, the proposed technique can be implemented through the frequency planning and DSP without any additional hardware parts. The results verifies that the proposed technique significantly increases the SNR of the targets by decreasing the noise floor due to the phase noise of the dominant leakage. Additionally, the proposed technique provides accurate distance information of the targets without deteriorating the target power.